\documentclass[prl,aps,superscriptaddress,twocolumn,floatfix,nofootinbib]{revtex4-2}

\usepackage{mathtools}
\usepackage[usenames,dvipsnames]{xcolor}
\usepackage[normalem]{ulem}
\usepackage{amsmath}
\usepackage{amssymb,mathrsfs}
\usepackage{graphicx}
\usepackage[colorlinks=true]{hyperref}
\newcommand{\sss}{\scriptscriptstyle}

\begin{document}

\title{Quantitative Analysis of Shock Wave Dynamics in a Fluid of Light}

\author{T. Bienaim\'e} \affiliation{Laboratoire
  Kastler Brossel, Sorbonne Universit\'e, CNRS, ENS-PSL Research
  University, Coll\`ege de France, Paris 75005, France}

\author{M. Isoard} \affiliation{Universit\'e Paris-Saclay, CNRS, LPTMS,
  91405, Orsay, France}\affiliation{Physikalisches Institut, Albert-Ludwigs-Universit\"at Freiburg, Hermann-Herder-Stra{\ss}e 3, D-79104 Freiburg, Germany} 

\author{Q. Fontaine}\affiliation{Laboratoire
  Kastler Brossel, Sorbonne Universit\'e, CNRS, ENS-PSL Research
  University, Coll\`ege de France, Paris 75005, France}

\author{A. Bramati}\affiliation{Laboratoire
  Kastler Brossel, Sorbonne Universit\'e, CNRS, ENS-PSL Research
  University, Coll\`ege de France, Paris 75005, France}

\author{A. M. Kamchatnov} \affiliation{Moscow Institute of
  Physics and Technology, Institutsky lane 9, Dolgoprudny, Moscow
  region, 141700, Russia} \affiliation{Institute of Spectroscopy,
  Russian Academy of Sciences, Troitsk, Moscow, 108840, Russia}

\author{Q. Glorieux}\affiliation{Laboratoire
  Kastler Brossel, Sorbonne Universit\'e, CNRS, ENS-PSL Research
  University, Coll\`ege de France, Paris 75005, France}

\author{N. Pavloff} \affiliation{Universit\'e Paris-Saclay, CNRS, LPTMS,
  91405, Orsay, France}

\begin{abstract}
We report on the formation of a dispersive shock wave in a nonlinear
optical medium. We monitor the evolution of the shock by tuning the incoming beam power.
The experimental observations for the position and 
intensity of the solitonic edge of the shock, as well as the location
of the nonlinear oscillations are well
described  by recent developments of Whitham modulation theory.
Our work constitutes a detailed and accurate benchmark for this approach. 
It opens exciting possibilities to engineer specific configurations of optical shock wave for studying wave-mean flow interaction.
\end{abstract}

\maketitle

In many different fields such as acoustics \cite{NL_acoustics}, plasma physics \cite{Jeffrey-Taniuti}, hydrodynamics \cite{StVenant1871,Bellevaux65,ovsyannikov_1979}, nonlinear optics \cite{Akhmanov1966}, ultracold quantum gases \cite{Kagan1996,Castin1996,Joseph2011,datta2020}, the short time propagation of slowly varying nonlinear pulses can be described discarding the effects of dispersion and dissipation.
The prototype of such an approach is given by the system of equations governing compressible gas dynamics \cite{Riemann}.
This type of treatment typically predicts that, due to nonlinearity, an initially smooth pulse steepens during its time evolution, eventually reaching a point of gradient catastrophe.
This is the wave-breaking phenomenon, which results in the formation of a shock wave \cite{Courant2015,Zeldovich-Raiser}. 
If, after wave breaking, dispersive effects dominate over viscosity, the shock eventually acquires a stationary nonlinear oscillating structure for which the width increases with diminishing dissipation \cite{Sagdeev1966}. 
In the case of weak dissipation the time for reaching a stationary regime can be quite long.
Gurevich and Pitaevskii \cite{Gurevich1973} made a major contribution to the field when they first realized the interest of studying the evolution of the associated dynamical structure, now called a dispersive shock wave (DSW).
Besides, they understood that a DSW can be described as a modulated nonlinear traveling wave and studied in the framework of the Whitham theory of modulations \cite{whitham_1965}.

In the present work we study the propagation of an optical beam in a nonlinear defocusing medium.
Wave-breaking and (spatial or temporal) dispersive shocks have already been observed in such a setting \cite{rothenberg_1989,couton_2004,wan_2007,Jia2007,ghofraniha_2007,barsi_2007,ghofraniha_2012,Elazar2012,Ghofraniha:12,fatome_2014,xu_2016,Wetzel2016,xu_2017,nuno_2019}.
However, all previous theoretical descriptions of experimental optical shocks either remained only qualitative or resorted to numerical simulations for reaching accurate descriptions.
Indeed, a realistic quantitative characterization of the experimental situation requires to take into account a number of nontrivial effects which sum up to a quite difficult task.
For instance, saturation effects, such as occurring in semiconductor doped glasses \cite{Coutaz:91} and in photorefractive media \cite{Kivshar-Agrawal}, can only be taken into account by using a nonintegrable nonlinear equation, even for a medium with a local nonlinearity.
Besides, both  ``Riemann invariants'' typically vary during the prebreaking period and this complicates the description of the nondispersive stage of the pulse spreading, even in a quasi unidimensional (1D) geometry.
Moreover, for realistic initial intensity pulse profiles, the post-breaking evolution corresponds, at best, to a so-called ``quasisimple'' dispersive shock \cite{Gurevich1987}, the characterization of which requires an elaborate extension of the Gurevich-Pitaevskii scheme. 
Finally, the nonintegrability of the wave equation significantly complicates the post-breaking description of the nonlinear oscillations within the shock. 
Despite these difficulties, it has recently become possible to combine several theoretical advances \cite{Gurevich1984,GKM_1989,KKE_1992,el_evolution_1993,el_general_1995,Kodama1999,El_2005,EGKKK_2007,forest_exact_2009,Kamchatnov_2019,IKP_2019,isoard-nondisp-2020} to obtain a comprehensive treatment of the nonlinear pulse spreading and the subsequent formation of a dispersive shock in a realistic setting \cite{isoard_2019,ivanov_2020}.
In this Letter, we provide a nonambiguous experimental evidence for the accuracy of this theory with a precise description of the main features of the shock.
This universal and quantitative benchmark is a major advance for manipulation and engineering of optical shockwaves.

\begin{figure*}[t!]
\centering
\includegraphics[width=0.8\textwidth]{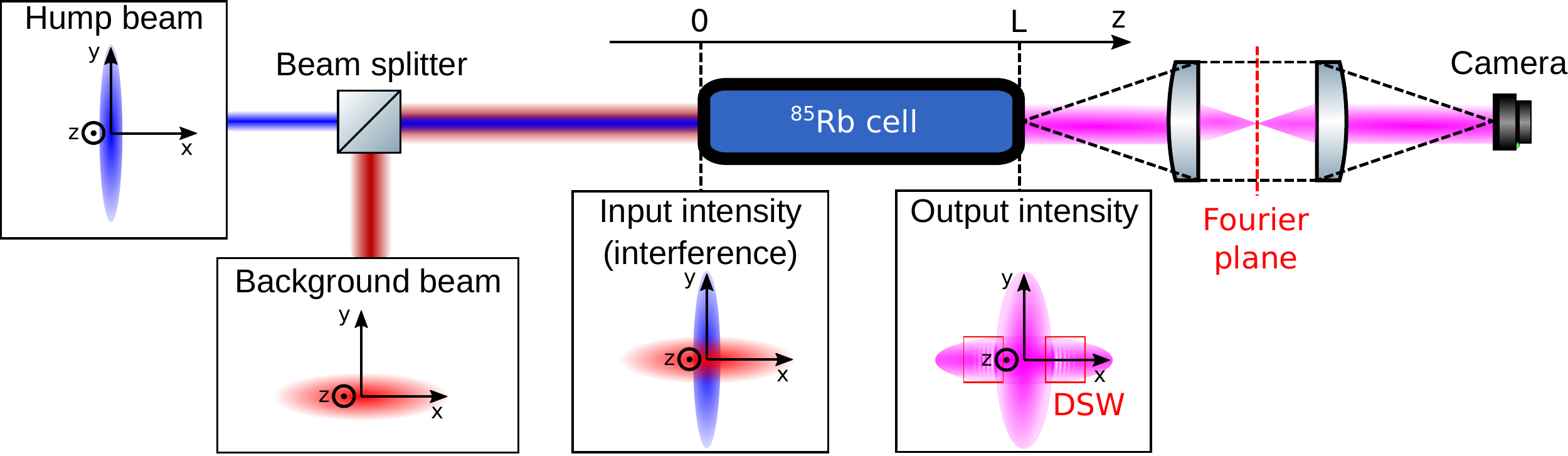}
\caption{Sketch of the experimental setup. To create the initial state we overlap the background and the hump beams on a beam splitter with their relative phase precisely adjusted such that they interfere constructively. This state then propagates inside the nonlinear medium consisting of a hot $^{85}$Rb vapor cell of length $L$. The insets represent cuts of the relevant intensity profiles in the plane perpendicular to the direction $z$ of propagation. The output intensity is recorded on a camera by direct imaging through two lenses in $4f$ configuration.  }\label{fig1}
\end{figure*}

We study the propagation of a laser field in a $L = 7.5$~cm-long cell filled with an isotopically pure $^{85}$Rb vapor (99\% purity)  warmed up to a controlled temperature of $120 \, ^\circ \text{C}$ to adjust the atomic vapor density.
We use a Ti:sapphire laser detuned by $- 3.9 \, \text{GHz}$ with respect to the $F = 3 \rightarrow F' $ transition of the D2-line of $^{85}$Rb at $\lambda_0 = 2 \pi / k_0 = 780 \, \text{nm}$.
For such a large detuning, the natural Lorentzian shape of the line dominates and the Doppler broadening $k_0 v \simeq 240 \, \text{MHz}$ can be safely neglected.
In these experimental conditions, the system is self defocusing (repulsive photon-photon interaction) and the transmission through the cell is $60 \, \%$. 
We find that this medium is well described by local photon-photon interactions, but contrary to  previous works \cite{PRL_PK_2018,michel_superfluid_2018,fontaine_2018,fontaine_2020,boughdad2019anisotropic}
we find it important to take into account the saturation of the nonlinearity to quantitatively describe the dynamics of the shock waves.

The input intensity profile is a cross-beam configuration of two vertically polarized laser beams, both propagating along the axis of the cell (denoted as O$z$), with their respective phase precisely adjusted such that the two beams interfere constructively, see Fig. \ref{fig1}.
One of the beams (which is denoted the \emph{hump}) is extended along the $y$ direction and significantly more intense than the other one (the \emph{background}) which is extended along the $x$ direction.
At the entrance of the cell ($z=0$) both beams have an  elliptic Gaussian profile.
The background beam has a power $P_0$ and waists $w_{x,0}>w_{y,0}$, whereas the hump has power $P_1$ and waists $w_{x,1}<w_{y,1}$ ; see Supplemental Material \cite{supp}.
During the initial nondiffractive stage of evolution, nonlinearity acts as an effective pressure which favors spreading of the hump in the $x$ direction along which is initially tighter collimated.
Conversely, the low intensity background experiences almost no spreading and behaves as a pedestal which triggers wave breaking of the hump during its spreading.
Each beam has a maximum entrance intensity ${\cal I}_{\alpha} = 2 P_{\alpha} / (\pi w_{x,\alpha} w_{y,\alpha} )$ ($\alpha=0$ or 1), and we explore the DSW dynamics for a fixed ratio ${\cal I}_{1} / {\cal I}_{0}$. 
We work in the deep nonlinear regime, with ${\cal I}_1= 20\,{\cal I}_0$.
This large value corresponds to a wave breaking distance typically shorter than the cell length, and makes it possible, by changing the total power $P_{\text{tot}}$ of the beams, to observe several stages of evolution of the DSW.
The total power can be increased up to $700 \, \text{mW}$ and is limited by the laser maximum output power.

\begin{figure*}[t!]
\centering
\includegraphics[width=\textwidth]{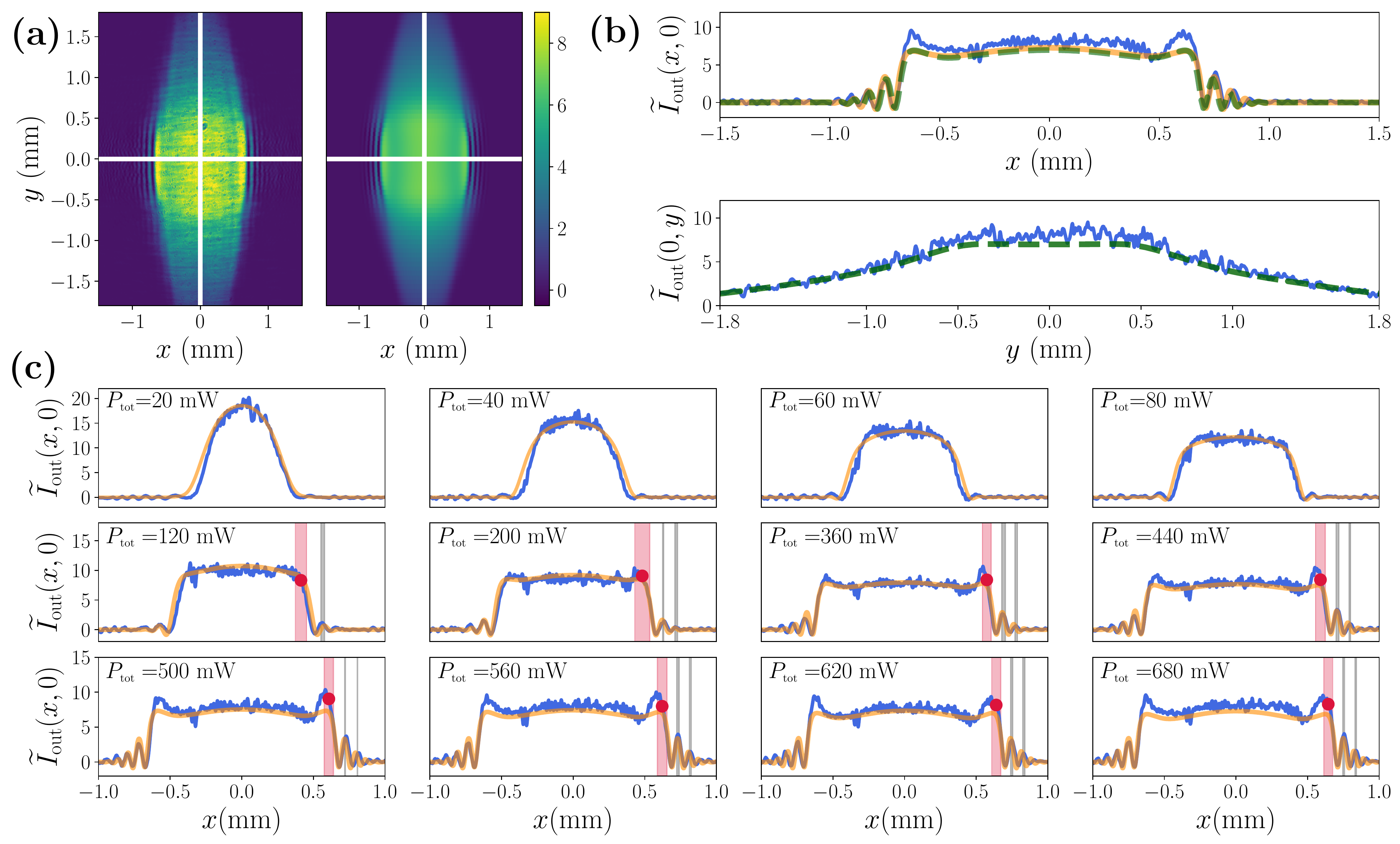}
\caption{(a) Left: experimental profile $\widetilde{I}_{\text{out}}$
  taken for $P_{\text{tot}} = 680 \, \text{mW}$. Right:
  Two-dimensional numerical simulations at the same total entrance
  power. (b) $x$ and $y$ profiles along
  the cuts represented by the two white lines on the two-dimensional
  profiles (a). The solid blue line represents the experimental data, the
  dashed green line the two-dimensional numerical simulation. On
  the $x$ profile, the orange line is a one-dimensional numerical
  simulation, from Eqs. \eqref{1D_saturable} and \eqref{a-init}.  (c)
  $\widetilde{I}_{\rm out}(x,0)$ for various total beam
  powers. The color code is the same as in (b). The vertical pink and
  gray bars on the right part of each intensity profile
  indicate the positions of the solitonic edge of the DSW and
  of the first maxima of oscillations within the DSW. The thickness of
  each bar represents the experimental uncertainty.}
\label{fig2}
\end{figure*}

We image the total field intensity $I_{\text{out}}(x,y)$ at the output of the cell on a camera.
In order to minimize the effect of absorption and increase the visibility of the DSW, we determine the \emph{normalized} output intensity
\begin{equation}\label{apodiz}
  \widetilde{I}_{\text{out}}(x,y) \equiv
  \frac{I_{\text{out}}(x,y) - I_{\text{out}}^{0}(x,y)   }
       { {\cal I}_{\text{out}}^{0} },
\end{equation}
where $I_{\text{out}}^{0}(x,y)$ is the intensity profile at the cell output when \emph{only} the background beam propagates through the medium (the hump beam is blocked). 
${\cal I}_{\text{out}}^{0}$ is the maximal value of $I_{\text{out}}^{0}(x,y)$.
$\widetilde{I}_{\text{out}}(x,y)$ is represented in Fig. \ref{fig2}(a).
Our theoretical description relies on only two parameters which characterize the photon-photon interaction, namely, the Kerr coefficient, $n_2$, and
the saturation intensity $I_{\text{sat}}$ [cf. Eq. \eqref{1D_saturable}]. Their values
$n_2 = 1.5\times 10^{-4}$ mm$^2$/W, and
$I_{\text{sat}} = 0.6$ W/mm$^2$
have been determined by comparing the experimental results with large-scale 2D
numerical simulations \cite{supp}.
The excellent agreement reached in Fig. 2 indicates that two effects -- saturable nonlinearity and linear absorption -- are the relevant physical ingredients for a theoretical description of our experiment.

In the regime $w_{1,x}\ll
w_{0,x}$ and ${\cal I}_1\gg {\cal I}_0$ we consider, the normalized output density  $\widetilde{I}_{\text{out}}$ becomes independent on the precise shape of the background beam. As a result, $\widetilde{I}_{\text{out}}(x,0)$ can be described by using a simplified 1D theoretical description, where a hump propagates over a background of {\it uniform} intensity ${\cal
  I}_0$.
Within the cell, the complex field amplitude at $y=0$, denoted as ${\cal A}(x,0,z)\equiv a(x,z)$, then obeys a 1D nonlinear Schr\"odinger equation where  the position $z$ along the axis of the beam plays the role of an effective ``time'' \cite{landau_vol8}.
 The equation, once included the nonlinearity saturation  and the linear
absorption \cite{Kivshar-Agrawal}, reads
\begin{equation}
  i \, \partial_z a = - \frac{1}{2 n_0 k_0} \partial^2_{x}  a
  + \frac{k_0\, n_2\, |a|^2}{1 + |a|^2/I_{\text{sat}}}
  \, a  -\frac{i}{\Lambda_{\rm abs}} \, a,
  \label{1D_saturable}
\end{equation}
where $n_0\simeq 1$ is the linear index of refraction and $\Lambda_{\rm abs}=30$ cm, which corresponds to a 60 \% transmission for a cell of length $L=7.5$ cm. 
The value of the effective amplitude at the entrance of the cell is taken as
\begin{equation}\label{a-init}
  a(x,0) =  \sqrt{{\cal I}_0}
  +  \sqrt{{\cal I}_{1}} \exp\left(- \frac{x^2}{w^2_{x,1}}\right),
\end{equation}
In order to evaluate the accuracy of the mapping to the 1D model of Eq. \eqref{1D_saturable}, we compare in the upper panel of Fig. \ref{fig2}(b) the corresponding value of $|a(x,L)|^2\exp(2L/\Lambda_{\rm abs})/{\cal I}_0 - 1$ with the experimental $\widetilde{I}_{\rm out}(x,0)$ and with the result of 2D simulations.
The excellent agreement is confirmed in Fig. \ref{fig2}(c) for the whole range of beam powers $P_{\rm tot}$.

The mapping to a 1D problem enables us to compare our measurements with recent analytical predictions.
In particular, if one neglects the linear absorption within the cell, 
for the initial intensity profile \eqref{a-init}, wave breaking occurs at a propagation distance \cite{isoard-nondisp-2020}
\begin{equation}\label{zwb}
  z_{\rm\sss WB}=4 \, \sqrt{\frac{n_0 I^*}{n_2}}\, 
    \frac{(1+I^*/I_{\rm sat})^2}{3+I^*/I_{\rm sat}}\cdot
    \frac{1}{{\rm max}\left|\frac{dI(x,0)}{dx}\right|},
\end{equation}
where $I(x,0)=|a(x,0)|^2$ is the entrance intensity and $I^*$ is the value $I(x^*,0)$ at point $x^*$ where $|dI(x,0)/dx|$ reaches its maximum. 
For low entrance power, no DSW is observed because $z_{\rm\sss WB}$ is larger than the cell length.
Wave breaking first occurs within the cell for a total power $P_{\rm\sss WB}$ such that $z_{\rm\sss WB}=L$. 
For our experimental parameters we obtain $P_{\rm\sss WB}=48$ mW. 
Numerical tests show that taking absorption into account does not modify notably this value.

\begin{figure}[t]
    \includegraphics[width=\linewidth]{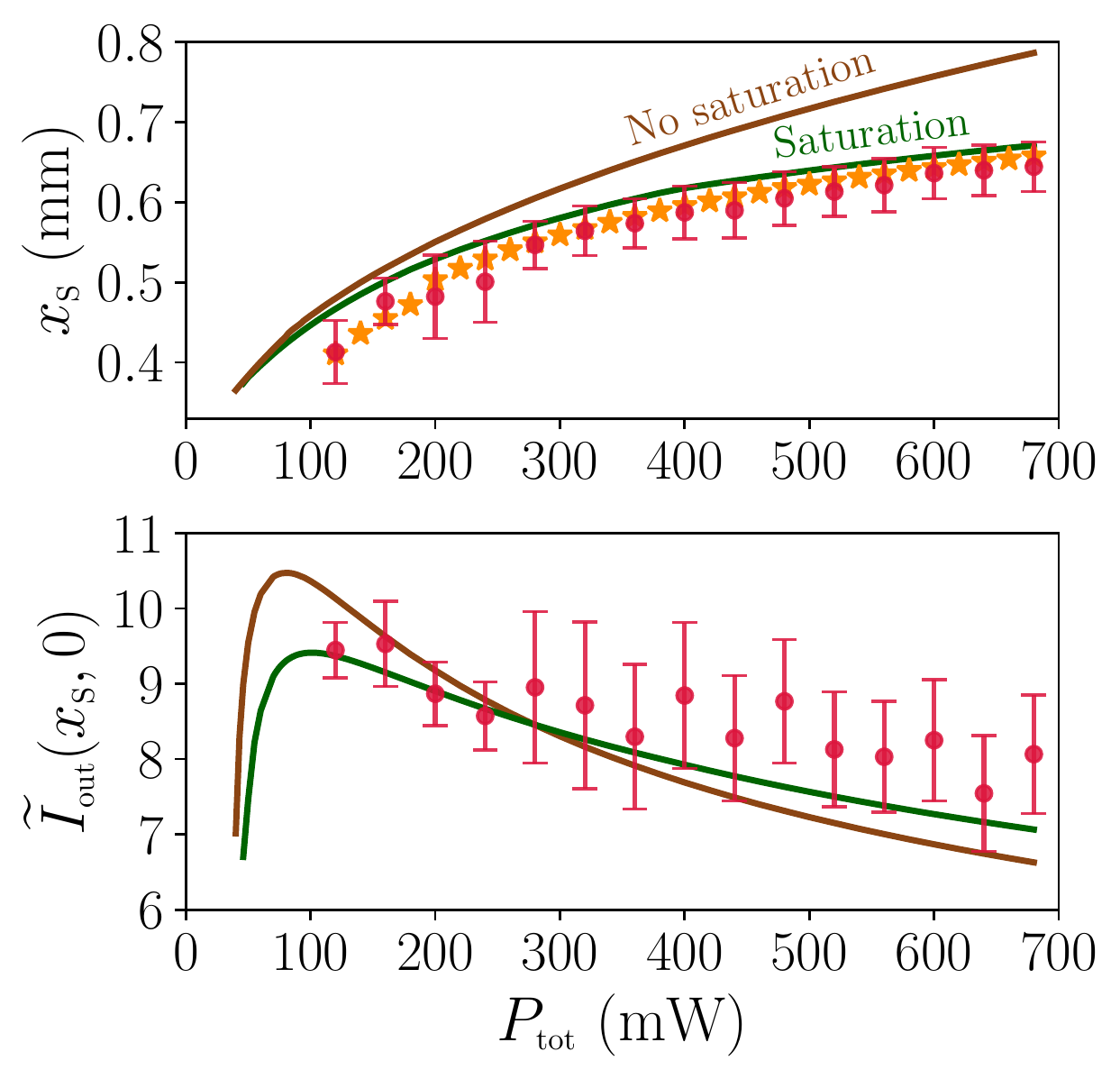}
    \caption{Characterization of the solitonic edge of the DSW as a function
      of the beam's power. The upper panel represents the position
      $x_s$ of the shock, and the lower one the corresponding
      intensity $\widetilde{I}_{\rm out}(x_s,0)$. In each panel, the
      red points with error bars are experimental results, from
      Fig. \ref{fig2}(c) and, in the upper one, the orange stars are the results of 1D numerical simulations of Eq. \eqref{1D_saturable}. The green solid line is the theoretical
      result, from Ref. \cite{ivanov_2020}. The brown solid line is
      the theoretical result in the absence of saturation.}
    \label{fig3}
\end{figure}

For a total power larger than $P_{\rm\sss WB}$, the DSW is formed and develops within the cell.
The physical phenomenon at the origin of the DSW is the following: large intensity perturbations propagate 
faster than small ones, so there exist values of $x$ reached at the same ``time'' by different intensities.
When this occurs first, the density gradient is infinite. 
This corresponds to the onset of a cusp catastrophe \cite{golubitsky_1978,Kamchatnov-book}, the nonlinear diffractive dressing of which is a dispersive shock wave.
This takes the form of a modulated oscillating pattern consisting asymptotically (i.e. at large $z$, or equivalently large $P_{\rm tot}$) in a train of solitons which,
away from the center of the beam, gradually evolves into a linear perturbation.
The position of its ``solitonic edge'' on the $y=0$ axis at the cell output  ($z=L$) is denoted as $x_s$.
It is located in Figs. \ref{fig2}(c) by a vertical red bar whose thickness represents the uncertainty 
on the estimation of $x_s$ from the experimental $\widetilde{I}_{\rm out}(x,0)$.
This uncertainty limits the experimental determination of $x_s$ to powers larger than 120 mW. 
The following maxima of oscillations, 
represented by vertical gray lines, are more 
precisely determined experimentally.
The technique devised in Ref. \cite{ivanov_2020} makes it possible to theoretically determine $x_s$ and the corresponding intensity $\widetilde{I}_{\rm out}(x_s,0)$.
As illustrated in Fig. \ref{fig3} the results of this analytical approach (green solid lines) compare well with the experimental data, although it does not take absorption into account.
Importantly, omission of the nonlinearity saturation  leads to incorrect results (brown solid line).

\begin{figure}[t]
  \includegraphics[width=\linewidth]{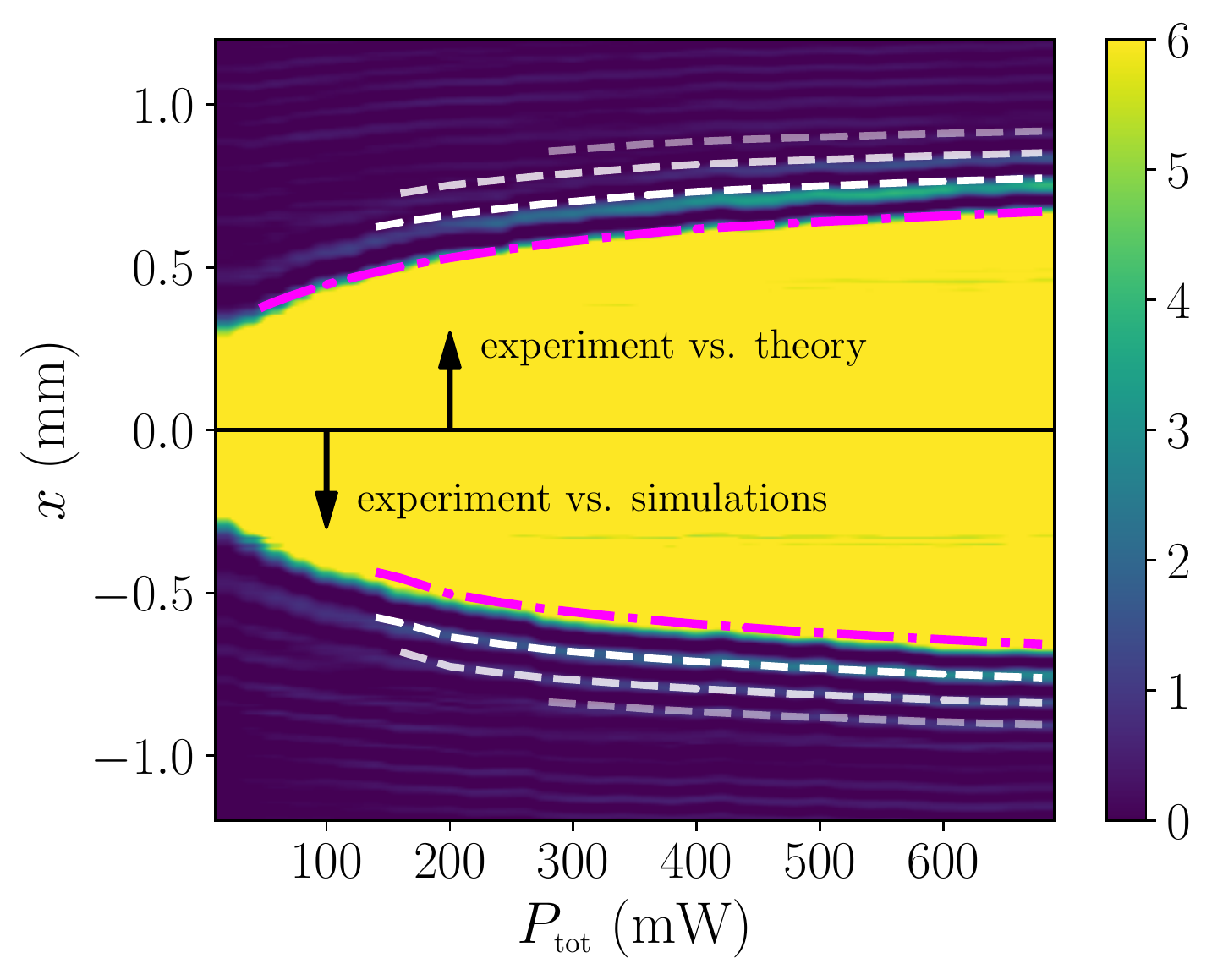}
  \caption{Color plot of the experimental intensity profiles
    $\widetilde{I}_{\text{out}}(x,0)$ as a function of
    $P_{\text{tot}}$. The purple dot-dashed line represents the edge
    $x_s$ of the DSW extracted from Whitham theory (upper part of the
    figure: $x>0$) and from 1D numerical simulations (lower part,
    $x<0$). In each half of the figure ($x\lessgtr 0$) the white
    dashed lines are the corresponding analytic predictions
    \eqref{max1} for the maxima of oscillation.}
    \label{fig4}
\end{figure}

One may study the DSW in an even more detailed way by locating the position of the maxima of the nonlinear oscillations.
While the
theoretical results for $x_s$ essentially rely on an approach due to El \cite{Gurevich1984,El_2005,Kamchatnov_2019} which is valid for any type of nonlinearity, the precise intensity profile within the DSW can be computed only for exactly integrable systems, i.e., by neglecting saturation effects.
The position-dependent oscillation period ${\cal L}(x,z)$ was computed in this framework in Ref. \cite{isoard_2019} for a parabolic initial intensity distribution.
Fitting the center of the intensity profile \eqref{a-init} by an inverted parabola, the positions $x_1$, $x_2$, and $x_3$ of the first maxima of oscillation of the DSW at the output of the cell are determined by
\begin{equation}\label{max1}
  x_1-x_s={\cal L}\left(\frac{x_s+x_1}{2},L\right),
\end{equation}
and by similar formulas obtained by replacing $x_1$ by $x_2$ (then $x_3$) and $x_s$ by $x_1$ (then $x_2$).
The results are compared with the experimental ones in the upper half ($x>0$) of Fig. \ref{fig4}.
The small offset in the position of the theoretical maxima with respect to the experimental ones observed in the figure is due to an initial small overshoot in the theoretical position of $x_s$ (cf. the green solid line in Fig. \ref{fig3}) which is itself due to the absence of absorption in the model.
Indeed, the 1D numerical simulations -- which do take absorption into account -- are in slightly better agreement with the experimental results for $x_s$ (cf. the orange stars in Fig. \ref{fig3}). Using the numerical $x_s$ in Eq. \eqref{max1} instead of the analytical one yields, for the maxima of oscillations, an excellent agreement with experiment, cf. the lower half of Fig. \ref{fig4}. Such a good agreement despite the fact that Eq. \eqref{max1} does not take saturation into account is not surprising: the rapid decrease of intensity away from the solitonic edge (cf. Fig. \ref{fig2}) significantly reduces the importance of saturation within the DSW.

It thus appears possible to give a detailed description of
precise experimental recordings of the intensity pattern of an optical
shock wave, not only thanks to numerical simulations, but on the basis of Whitham's modulation
theory. This is an important validation of recent
advances in this approach, which is no longer restricted to integrable
systems or idealized initial configurations. We are reaching a point
where these progresses make it possible not only to study DSWs {\it
  per se}, but also as tools for prospecting new physical
phenomena, such as the type of wave-mean flow interaction recently identified in Ref. \cite{CEH_2019}:  our platform is ideally designed to investigate scattering of elementary excitations by a DSW, a study which is also relevant to the domain of analogue gravity. Indeed, as discussed in the Supplemental Material \cite{supp}, a dispersive shock can be considered as
an exotic model of acoustic white hole, and the good
experimental control and theoretical
understanding of this structure demonstrated in the present work opens
the prospect of a detailed investigation of the corresponding induced background fluctuations.

\end{document}


\title{\huge Supplemental Material}

\maketitle

\setcounter{figure}{0}
\renewcommand{\theequation}{S\arabic{equation}}
\renewcommand{\thefigure}{S\arabic{figure}}
\renewcommand{\thesection}{S\arabic{section}}
\renewcommand{\thesubsection}{S\arabic{subsection}}
\setcounter{equation}{0} 

\section{Two-dimensional numerical simulations}

\begin{figure*}[t]
    \includegraphics[width=1\linewidth]{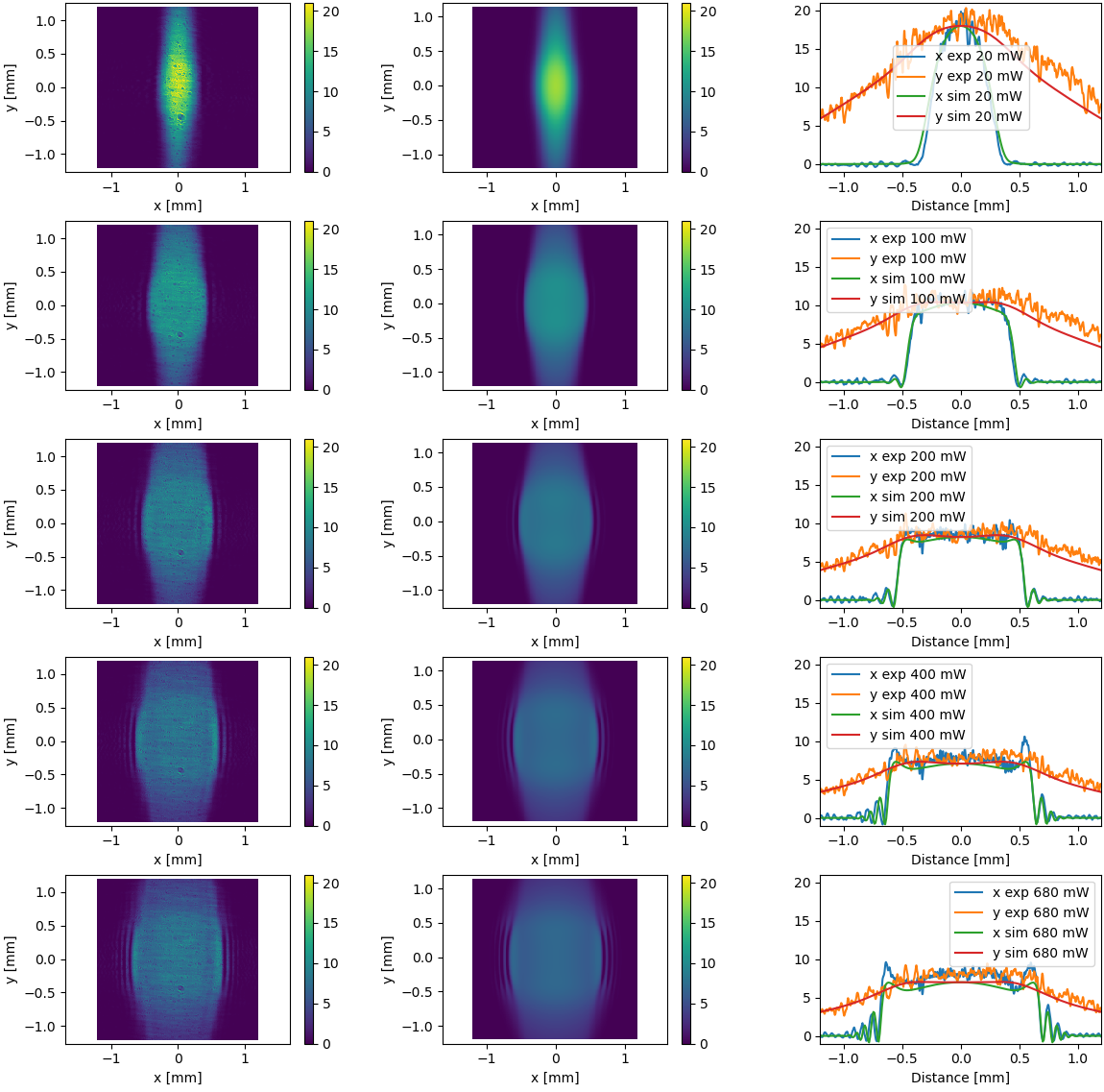}
    \caption{Normalized intensity $\widetilde{I}_{\text{out}}(x,y)$
      for the experimental data and the two-dimensional numerical
      simulations for $P_{\text{tot}} = 20, 100, 200, 400$ and $680 \, \text{mW}$ from top to bottom row. The first column shows
      the experimental data. The second column is the corresponding
      numerical simulation with the parameters used in the text. The third column compares, for each total incoming power,
      $\widetilde{I}_{\text{out}}(x,0)$ and
      $\widetilde{I}_{\text{out}}(0,y)$ for the experimental data and
      the simulations.}
    \label{figS2}
\end{figure*}

The amplitudes of the incoming beams are described by the complex fields $\mathcal A_{\alpha}(x,y,z=0)$ where $\alpha=0$ for the background and $\alpha=1$ for the hump. Their explicit expressions read
\begin{equation}\label{s1:eq1}
  \mathcal A_{\alpha}(x,y,z=0) = \sqrt{{\cal I}_{\alpha}}
  \exp \left( - \frac{x^2}{w^2_{x,\alpha}} -
  \frac{y^2}{w^2_{y,\alpha}} \right)\; ,
\end{equation}
where the values of the different waists are 
$w_{x,0}= 3.35 \, \text{mm}$, $w_{y,0} = 0.4 \, \text{mm} $, $w_{x,1} = 0.23 \, \text{mm}$ and $w_{y,1} = 1.65 \, \text{mm}$.  The
total field at the entrance of the cell is then given by $\mathcal
A(x,y,0) = \mathcal A_{0}(x,y,0) + \mathcal A_{1}(x,y,0)$; it corresponds to a total incident power
$P_{\text{tot}} = \int \mathrm dx \,
\mathrm dy \, |\mathcal A(x,y,0)|^2$. From expressions \eqref{s1:eq1} one obtains
\begin{equation*}
  \frac{P_{\text{tot}}}{P_{0} } = 1
  + \beta \frac{w_{x,1}   w_{y,1}}{w_{x,0} w_{y,0}} + 4
  \sqrt{\beta \, \frac{w_{x,1}   w_{y,1}}{w_{x,0} w_{y,0}}}  ,
\end{equation*}
where $P_0= \int \mathrm dx \,
\mathrm dy \, |\mathcal A_0(x,y,0)|^2 = \tfrac{\pi}{2}\mathcal I_0 w_{x,0} w_{y,0}$ is the incoming power of the background beam and $\beta={\cal I}_1/{\cal I}_0$. In all this work we impose $\beta=20$ which corresponds to a highly nonlinear regime and favors a rapid formation of the dispersive shock wave (DSW).
Within the nonlinear cell, in the paraxial approximation, the amplitude of the beam obeys a 2D 
generalized nonlinear Schr\"odinger equation:
\begin{equation}
  i \partial_z \mathcal A
  = - \frac{1}{2 n_0 k_0} \vec{\nabla}^2_{\perp} \mathcal A
  + \frac{k_0 n_2|\mathcal A|^2}{1 + |\mathcal A|^2/I_{\text{sat}}}\mathcal A
  -\frac{i}{\Lambda_{\rm abs}}
  \mathcal A , \label{2D_saturable}
\end{equation}
where $\vec{\nabla}^2_{\perp}=\partial_x^2+\partial_y^2$, $n_2$ is the nonlinear Kerr coefficient, $I_{\rm sat}$ the saturating density of the nonlinearity and $\Lambda_{\rm abs}$ describes the effects of absorption. The power at the exit of our cell of length $L$ is only 60 \% of the incoming one, this corresponds to $\Lambda=4 L$ (i.e., $\exp(-2 L/\Lambda_{\rm abs})=0.6$).
Changing the value of the total incoming power $P_{\rm tot}$ modifies the position $z_{\rm\sss WB}$ at which wave breaking  
occurs and makes it possible to observe different stages of evolution of the DSW at the output $z=L$ of the nonlinear cell, see the left column of Fig. \ref{figS2}.
We solved Eq. \eqref{2D_saturable} by two independent numerical methods (the split step method and the fourth order Runge-Kutta technique) for different values of $P_{\rm tot}$. The best agreement with the experimental profiles is obtained for  the choice of parameters $n_2 = 1.5\times 10^{-4}$ mm$^2$/W and
$I_{\text{sat}} = 0.6$ W/mm$^2$.
Fig. \ref{figS2} compares the normalized intensity profiles obtained
experimentally and numerically. It reveals very good quantitative
agreement between the two.

\begin{figure}[t]
    \includegraphics[width=\linewidth]{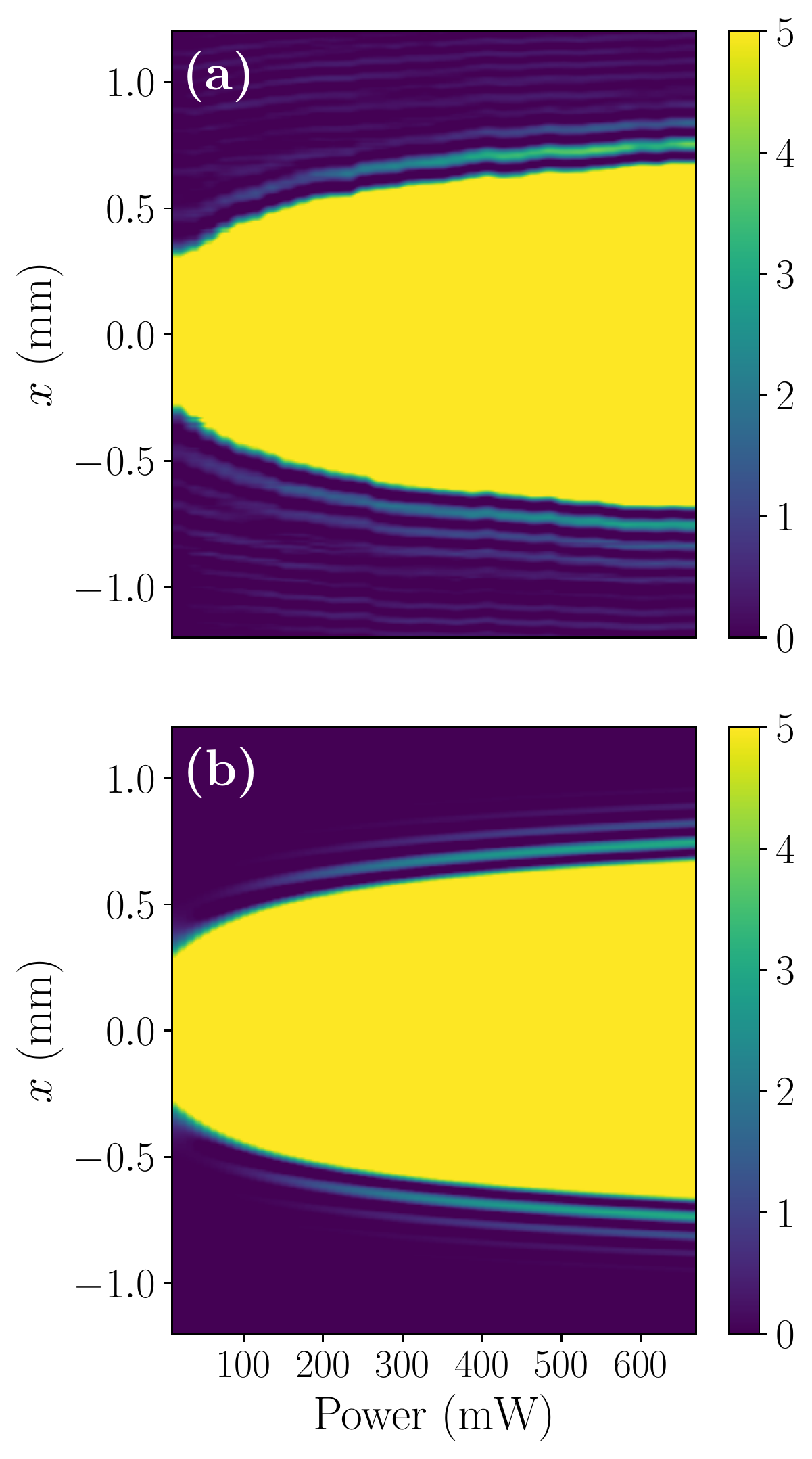}
    \caption{Normalized intensity $\widetilde{I}_{\text{out}}(x,y=0)$
      as a function of $P_{\text{tot}}$. (a) Experimental
      data. (b) Two-dimensional numerical simulations of Eq. \eqref{2D_saturable} .}
    \label{figS4}
\end{figure}

Another test of the relevance, over all the range of input powers, of our choice of parameters
is the comparison of the experimental result for the normalized output intensity $\widetilde{I}_{\text{out}}(x,y=0)$ with the ones of numerical simulations. This check is performed in Fig. \ref{figS4}, in which one can observe an overall agreement for the front $x_s$ of the shock (interface between the dark and yellow regions) and also for the location of the first maxima of oscillation whithin the dispersive shock wave. The agreement is very good for all incident powers (from 20 mW up to 680 mW).

\section{A dispersive shock wave considered as a white hole}

\begin{figure}
    \centering
    \includegraphics[width=\linewidth]{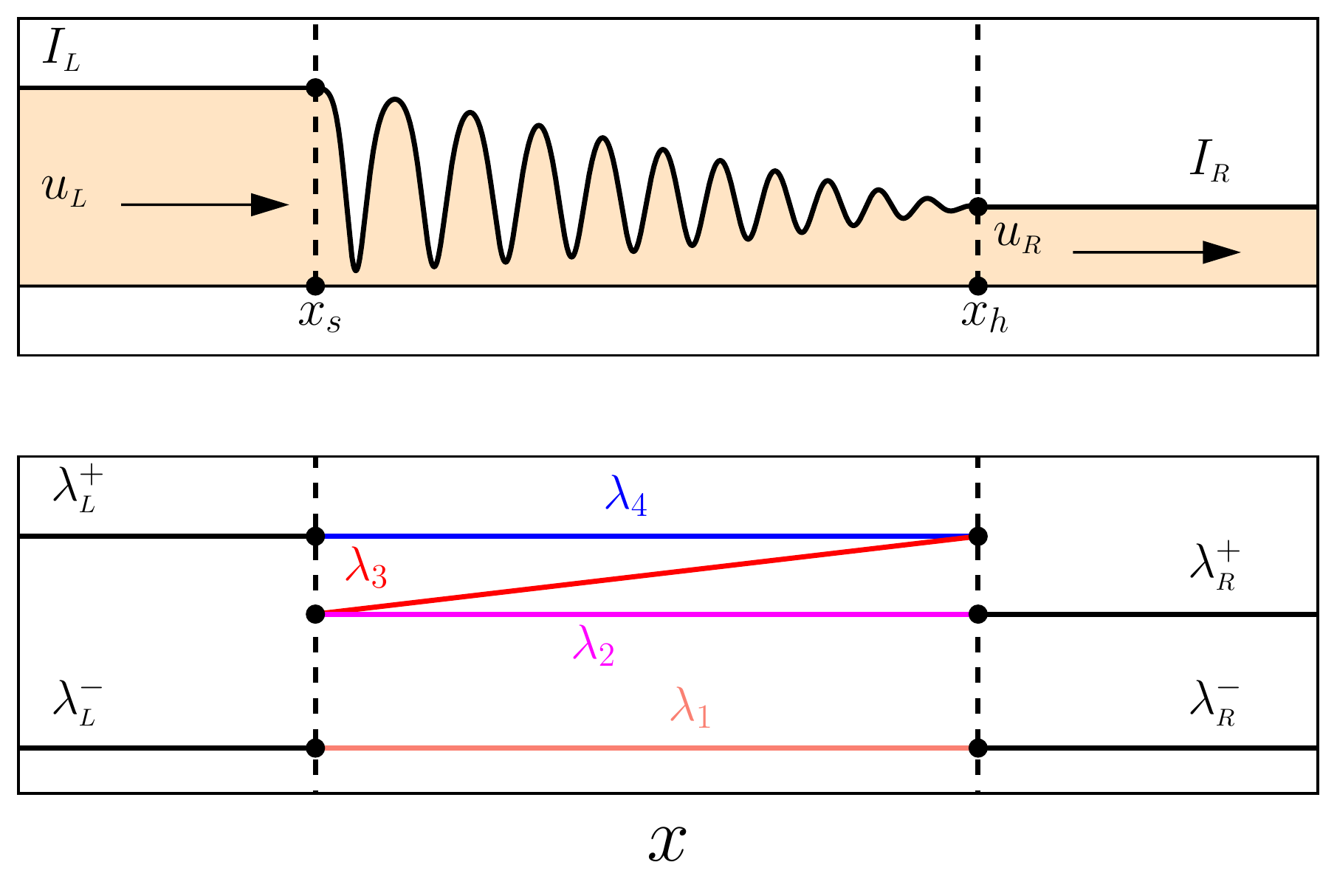}
    \caption{Upper plot: schematic representation of the intensity
      pattern $I(x,t)$ in the presence of a Gurevitch-Pitaevskii DSW. Lower
      plot: corresponding arrangement of the Riemann invariants.}
    \label{fig:GP_DSW}
\end{figure}

In this section, we argue that a dispersive shock wave realizes a
transsonic flow which can be considered as an implementation of an
analog white hole. For simplying the presentation, we do not take into
account absorption or saturation of nonlinearity: this
amounts to take $I_{\rm sat}=\Lambda_{\rm sat}=\infty$ in a 1D version of 
Eq. \eqref{2D_saturable}. Also, we do not discuss the complicated
situation considered in the main text, but rather the ideal case of a
perfect Gurevitch-Pitaevskii configuration issued from Riemann initial
conditions, with piece-wise initial intensity and velocity,
discontinuous at $x=0$, see e.g., Ref. \cite{EL1995}. 
In such a
configuration
the intensity and the Riemann invariants behave as sketched in
Fig. \ref{fig:GP_DSW}, where the dispersive shock propagates to the
right, with asymptotic intensity and velocity\footnote{Strictly
  speaking the use of the term "velocity" is improper. Its use amounts
  to consider the $z$ variable as an effective time, as done in the
  main text.} $I_{\sss R}$ and $u_{\sss R}$ at the right and
$I_{\sss L}$ and $u_{\sss L}$ at the left ($I_{\sss L}>I_{\sss R}$ and
$u_{\sss L}>u_{\sss R}$).
Outside of the shock region, the dispersionless Riemann invariants are
$\lambda_{\sss L/R}^{\pm}=\tfrac12 u_{\sss R/L}\pm c_{\sss R/L}$ where
$c_{\sss R/L}=\sqrt{n_2 I_{\sss R/L}/n_0}$ is the speed of sound in a
uniform fluid with intensity $I_{\sss R/L}$. Within the shock, only
one of the Whitham-Riemann invariants (the colored $\lambda_i$'s of
Fig. \ref{fig:GP_DSW}) is not constant: $\lambda_3$. Its value is
determined by \cite{Kamchatnov-book}
\begin{equation}
x/t=v_3(\lambda_1,\lambda_2,\lambda_3,\lambda_4),
\end{equation}
with $\lambda_1=\lambda^-_{\sss R}= \lambda^-_{\sss L}$,
$\lambda_2=\lambda^+_{\sss R}$, $\lambda_4=\lambda^+_{\sss L}$ and
\begin{equation}
  v_3=\tfrac12\sum_{i=1}^4\lambda_i-
  \frac{(\lambda_4-\lambda_3)(\lambda_3-\lambda_2)\mathrm{K}(m)}
    {(\lambda_3-\lambda_2)\mathrm{K}(m)-(\lambda_4-\lambda_2)\mathrm{E}(m)},
\end{equation}
where 
\begin{equation}\label{4-41}
    m=
\frac{(\lambda_2-\lambda_1)(\lambda_4-\lambda_3)}
{(\lambda_4-\lambda_2)(\lambda_3-\lambda_1)}\; ,
\end{equation}
and $\mathrm{K}(m)$ and $\mathrm{E}(m)$ are the complete
elliptic integrals of the first and second kind, respectively.

At the location $x=x_s$ of the solitonic edge, $\lambda_2=\lambda_3$ and $v_3$ takes the
limiting value $v_3=\tfrac12(\lambda_1+2\lambda_2+\lambda_4)$ which is
also the velocity $v_s$ of the solitonic edge. This yields
\begin{equation}
    v_s=u_{\sss R}+c_{\sss L}\; .
\end{equation}
At the opposite ``harmonic edge'' (of position $x_h$) one has instead
$\lambda_3=\lambda_4$ and
$v_3=2\lambda_4+\tfrac12(\lambda_2-\lambda_1)^2/(\lambda_1+\lambda_2-2\lambda_4)$. This
yields for the velocity of the harmonic edge
\begin{equation}
    v_h=u_{\sss L}+2c_{\sss L}-\frac{c^2_{\sss R}}{2c_{\sss L}-c_{\sss R}}.
\end{equation}
In the reference frame ${\cal R}'$ in which the solitonic edge is at
rest, the fluid of light at the right boundary ($x\to+\infty$)
propagates at velocity $u'_{\sss R}=u_{\sss R}-v_s=-c_{\sss L}$: the flow is
directed to the left and is supersonic (clearly
$c_{\sss L}>c_{\sss R}$ since $I_{\sss L}>I_{\sss R}$). On the other
hand, still in ${\cal R}'$, the fluid at the left boundary has a
velocity $u'_{\sss L}=u_{\sss L}-v_s=c_{\sss L}-2 c_{\sss R}$. One can easily
verify that this quantity is lower in norm than the left speed of
sound $c_{\sss L}$.  If $I_{\sss L}\ge 4 I_{\sss R}$, the DSW admits a
vacuum point \cite{EL1995} and the velocity $u'_{\sss L}$ is positive.  This velocity is directed to the left only if
$I_{\sss L}< 4 I_{\sss R}$, i.e., if the mismatch in asymptotic
intensities is not too large. In this case the flow in ${\cal R}'$ is always directed to the left, and is upstream supersonic and downstream subsonic. This is an analogue white hole, with the important proviso
that the structure is not stationary: $I$ and $u$ are time-dependent
in any reference frame.  However, it has been established in the case
of Korteweg-de Vries equation \cite{Gurevich1987b,Avilov87} and also
for the nonlinear Schr\"odinger equation
\cite{Kamchatnov2004,Larre2012b} that if one takes dissipation into
account, after a (typically long) transient, the structure becomes
stationary and the analogy with a white hole is then more
straightforward. 

Note however, that even in a stationary case, as for any dispersive
analog, the horizon is ``fuzzy'', i.e., cannot be determined by the
condition that the flow velocity is equal to the local sound velocity,
because within the DSW the density varies over too short a distance to
enable a proper definition of a local sound velocity. Actually, in the
domain of analogue gravity, it is more important to correctly describe
the scattering of small amplitude linear waves than to try to
precisely locate the horizon. For instance the correct Hawking
temperature of a dispersive analogue is determined by the low energy
behavior of the scattering coefficients, and not by the
characteristics of the flow at an elusive horizon. Pursuing this line
of reasoning to its ultimate consequence leads to consider analogue
gravity as a sub-domain of wave-mean flow interaction. In this line,
it seems legitimate to follow the approach of Ref. \cite{CEH_2019}
which studies the scattering of linear waves by replacing the DSW by
an average mean flow. The good results obtained thanks to this
method suggest a possible solution of the above discussed difficulty
for locating the horizon of a DSW: a proper definition could be
reached by studying the corresponding average mean flow.